\documentclass[sn-mathphys-num]{sn-jnl}

\usepackage{graphicx}%
\usepackage{multirow}%
\usepackage{amsmath,amssymb,amsfonts}%
\usepackage{amsthm}%
\usepackage{mathrsfs}%
\usepackage[title]{appendix}%
\usepackage{xcolor}%
\usepackage{textcomp}%
\usepackage{manyfoot}%
\usepackage{booktabs}%
\usepackage{algorithm}%
\usepackage{algorithmicx}%
\usepackage{algpseudocode}%
\usepackage{listings}%
\usepackage{float}
\usepackage{enumitem}
\usepackage{lmodern}
\usepackage{anyfontsize}
\usepackage{caption}
\usepackage{cleveref}

\theoremstyle{thmstyleone}%
\theoremstyle{thmstyletwo}%

\theoremstyle{thmstylethree}%

\begin{document}

\title[An Application of Membrane Computing to Humanitarian Relief via Generalized Nash Equilibrium]{An Application of Membrane Computing to Humanitarian Relief via Generalized Nash Equilibrium}


\author*[1]{\fnm{Alejandro} \sur{Luque-Cerpa}}\email{luque@chalmers.se}

\author[2,3]{\fnm{David} \sur{Orellana-Martín}}\email{dorellana@us.es}

\author[3]{\fnm{Miguel Á.} \sur{Gutiérrez-Naranjo}}\email{magutier@us.es}

\affil*[1]{\orgdiv{Department of Computer Science and Engineering}, \orgname{Chalmers University of Technology}, \orgaddress{\country{Sweden}}}

\affil[2]{\orgdiv{Research Group of Natural Computing}, \orgname{Universidad de Sevilla}, \orgaddress{\country{Spain}}}

\affil[3]{\orgdiv{Department of Computer Science and Artificial Intelligence}, \orgname{Universidad de Sevilla}, \orgaddress{\country{Spain}}}


\abstract{Natural and political disasters, including earthquakes, hurricanes, and tsunamis, but also migration and refugees crisis, need quick and coordinated responses in order to support vulnerable populations. In such disasters, nongovernmental organizations compete with each other for financial donations, while people who need assistance suffer a lack of coordination, congestion in terms of logistics, and duplication of services. From a theoretical point of view, this problem can be formalized as a Generalized Nash Equilibrium (GNE) problem. This is a generalization of the Nash equilibrium problem, where the agents' strategies are not fixed but depend on the other agents' strategies. In this paper, we show that Membrane Computing can model humanitarian relief as a GNE problem. We propose a family of P systems that compute GNE in this context, and we illustrate their capabilities with Hurricane Katrina in 2005 as a case study.}




\maketitle

\section{Introduction}
According to \cite{DataPage2023}, the economic cost of damages as a result of global natural disasters in 2023 was 203.35 billion of USA dollars. They include droughts, floods, extreme weather, extreme temperatures, landslides, dry mass movements, wildfires, volcanic activity, and earthquakes. Not only natural but also political disasters, such as migration and refugee crises, need quick and coordinated responses to support vulnerable populations. In such disasters, nongovernmental organizations (NGOs) are the main actors to reduce suffering and mortality and support quality of life. However, several studies (e.g. \cite{Kopinak2013,NagurneyMain}) point out that humanitarian aid has not been successful due to a lack of coordination, congestion in terms of logistics, and duplication of services. 

Although there are huge differences between humanitarian logistics and commercial logistics \cite{Ortuño2013}, both problems can be formalized in a similar way. In \cite{DBLP:journals/anor/ToyasakiW14}, the authors compare fundraising with and without an earmarking option using optimization models. It seems to be the first study where game theory is considered in order to model the interaction between donors and humanitarian organizations.

In \cite{NagurneyMain}, the authors develop a Generalized Nash Equilibrium network model for post-disaster humanitarian relief by nongovernmental organizations which is the starting point for our study by using Membrane Computing techniques. In such a paper, the authors consider Hurricane Katrina as a case study, the costliest disaster in the history of the United States. This hurricane caused huge damage to property and infrastructure, left 450,000 people homeless, and took 1833 lives in Florida, Texas, Mississippi, Alabama, and Louisiana. For this work, we take the data of the disaster provided by Nagurney {\it et al.} \cite{NagurneyMain}.

From a theoretical point of view, this problem can be formalized as a Generalized Nash Equilibrium (GNE) problem. This is a generalization of the Nash equilibrium problem, where the agents' strategies are not fixed but they depend on the other agents' strategies and it can be considered as a problem in the area of Evolutionary Game Theory (EGT) \cite{EGT}. Beyond the intrinsic interest of GNE as a theoretical problem, it can be applied to a wide range of applications, such as the control of interacting vehicles \cite{BRITZELMEIER202015146}, the intersection management problem \cite{doi:10.1080/02331934.2020.1786088} or the energy market \cite{ DBLP:conf/eucc/Martinez-Piazuelo22}. GNE (or its simplified version, Nash equilibrium) has also provided a theoretical model for other helping actions such as blood donations \cite{NAGURNEY2019103,NagurneyMain}, competition for medical supplies \cite{Nagurney2021}, or logistics for humanitarian services \cite{Nagurney2022,NAGURNEY2019212} among many others.

Membrane Computing is a well-known bio-inspired computing paradigm \cite{DBLP:journals/jcss/Paun00,HandbookMC10}, whose devices, called P systems, have been used to successfully model many real-life problems, such as the dynamics of the population of giant panda in captivity \cite{DBLP:journals/nc/RongDVZQP23}, fault propagation paths in power systems \cite{DBLP:journals/access/WangWHWPPV19} or the ecosystem of some scavenger birds \cite{DBLP:conf/membrane/CardonaCMPPS09} among many others. To the best of our knowledge, the first time that a GNE problem was simulated by Membrane Computing devices was in Luque-Cerpa {\it et al.} \cite{LuqueCerpa2024}. This paper follows the research line started in that paper by showing that the Membrane Computing paradigm can be useful in the simulation of humanitarian relief. In opposition to that paper, we propose a family of P systems that computes GNE on a game out of the framework of Evolutionary Game Theory.

The paper is organized as follows: Section \ref{GNE_model} recalls some theoretical aspects related to how GNE can be used in the framework of humanitarian relief after a disaster. Section \ref{Transition} gives some basic information about the P system model used for the simulation. In Section \ref{section:design}, the design of our Membrane Computing device for dealing with humanitarian relief based on the GNE problem is presented. Next, Section \ref{section:experiments} shows, as a case study, the use of our device for simulation of humanitarian relief with the data of Hurricane Katrina, and finally, the paper ends with some conclusions and some future research lines.

\section{GNE model for post-disaster humanitarian relief}\label{GNE_model}
The Generalized Nash Equilibrium Problem (GNEP), first presented by G. Debreu \cite{Debreu1952} in 1952, is a generalization of the Nash Equilibrium problem in which the players' strategies depend on their rivals' strategies. Constraints in the game define these dependencies, and the payoffs obtained by each player depend on the strategies selected. Each player can choose only one strategy at a time, although this choice can be changed for subsequent actions. The guiding idea here is that players typically select courses of action that provide them with greater rewards, but the selected strategy may vary over time to adapt to the behavior of the rest of the players. When no player in this situation may change its strategy and whereas other agents keep using their current ones, a Nash equilibrium is reached \cite{Nash1951}.

Generalized Nash Equilibrium problem can be concretized into problems from many different areas. Among other applications, in~\cite{NagurneyMain} authors model the humanitarian relief post-disaster problem as a GNEP. This problem is then transformed into an optimization problem. The formulation is as follows.



Let us have $m$ NGOs that provide disaster relief to $n$ different locations. NGOs compete to provide relief items to demand points. Let $q_{ij} \geq 0$ be the amount of items that the NGO $i$ provides to the demand point $j$. Let $s_i$ be the maximum amount of relief items that the NGO $i$ can provide. Then, the following equation must hold for each NGO $i$: $\sum\limits_{j=1}^n q_{ij} \leq s_i$; that is, the total of the items provided by each NGO cannot be higher than the maximum relief items that such an NGO can provide.

Furthermore, for each NGO $i$ and each location $j$, we can consider the mapping $c_{ij}: \mathbb{R}^+ \rightarrow \mathbb{R}^+$, which maps $q_{ij}$ to $c_{ij}(q_{ij})$; that is, the mapping $c_{ij}$ represents the cost of sending the relief to the location $j$ by the NGO $i$.

Similarly, each NGO $i$ gets utility associated with the relief items it sent to the location $j$. The utility over all demand points is measured by $\sum\limits_{j=1}^n \gamma_{ij} q_{ij}$, where $\gamma_{ij}$ is a positive factor. Each NGO $i$ has a positive weight $\omega_i$ associated with such a utility measure that represents the monetization of this utility, and the monetization of an NGO $i$ is measured by $\omega_i \sum\limits_{j=1}^n \gamma_{ij} q_{ij}$. Finally, each NGO $i$ receives funding from donors based on media attention and the visibility of NGOs at location $j$. These funds are represented by $\beta_i \sum\limits_{j=1}^n P_j(q)$, where $q$ is the vector of all the item flows of all the NGOs to all the demand points, $P_j(q)$ represents the funds in donation dollars due to the visibility of all NGOs at location $j$, and $\beta_i$ represents the proportion of total donations collected that is received by NGO $i$. 

Other constraints are usually imposed by an authority. For example, there are lower ($\underline{d}_j$) and higher bounds ($\overline{d}_j$) for the number of relief items needed at a location $j$. These constraints are expressed as $\sum\limits_{i=1}^m q_{ij} \geq \underline{d}_j$ and $\sum\limits_{i=1}^m q_{ij} \leq \overline{d}_j$, which limit the amount of objects distributed by all NGOs to a specific location $j$. Another important assumption is that NGOs have enough resources to comply with the lower bounds of relief items required by all locations, that is, $\sum\limits_{i=1}^m s_i \geq \sum\limits_{j=1}^n \underline{d}_j$.
Finally, the optimization problem faced by the NGO $i$ can be expressed as

$$ \text{Minimize } \ -\beta_i  \sum\limits_{j=1}^n P_j(q) - \omega_i \sum\limits_{j=1}^n \gamma_{ij} q_{ij} + \sum\limits_{j=1}^n c_{ij}(q_{ij})$$

\noindent which, subject to the constraints indicated, is equivalent (Theorem 1, \cite{NagurneyMain}) to the following optimization problem:

\begin{equation} 
    \text{Minimize } \ - \sum\limits_{j=1}^n P_j(q) -  \sum\limits_{i=1}^m \sum\limits_{j=1}^n \frac{\omega_i \gamma_{ij} }{\beta_i} q_{ij} +  \sum\limits_{i=1}^m \sum\limits_{j=1}^n \frac{1}{\beta_i} c_{ij}(q_{ij}) \label{problem}
\end{equation}  

\textbf{Note} \textit{(Existence and uniqueness)} A solution $q^*$ to the optimization problem (\ref{problem}) is guaranteed to exist when the objective function consists of continuous functions and the feasible set defined by the constraints is compact. In \cite{NagurneyMain}, the functions employed satisfy these conditions, and because the objective function is strictly convex, the uniqueness of the solution is also guaranteed. 

\subsection{Euler method applied to the game theory model} \label{sectioneuler}

To solve the minimization problem (\ref{problem}), we can use a variational inequality formulation of the problem\footnote{An interested reader can find a detailed description in \cite{NagurneyMain}.} and then apply the Euler method \cite{dupuis1993dynamical, nagurney2012projected} to obtain closed-form expressions in the $q_{ij}$ and Lagrange multipliers $\lambda_i, \lambda_j^1, \lambda_j^2$ associated with the constraints. Such Euler method approximates the solution iteratively. We have the following expressions of $q_{kl}^{t+1}$ at each iteration $t$ for $k=1,...,m$; $l=1,...,n$:

\begin{equation}\label{eq1}
    q_{kl}^{t+1} = \max \Big\{0, \big\{ q_{kl}^t + a_t \Big( \sum\limits_{l=1}^n \Big( \frac{\partial P_j(q^t)}{\partial q_{kl}} \Big) + \frac{\omega_k \gamma_{kl}}{\beta_k} - \frac{1}{\beta_k} \frac{\partial c_{kl}(q_{kl}^t)}{\partial q_{kl}} - \lambda_k^t + {\lambda_l^1}^t - {\lambda_l^2}^t \Big) \} \Big\}
\end{equation}

\begin{equation}\label{eq2}\lambda_k^t = \max \Big\{0, \lambda_k^t + a_t \Big( -s_k + \sum\limits_{l=1}^n q_{kl}^t \Big) \Big\}
\end{equation}

\begin{equation}\label{eq3}{\lambda_l^1}^{t+1} = \max \Big\{0, {\lambda_l^1}^t + a_t \Big( - \sum\limits_{k=1}^m q_{kl}^t + \underline{d_l}  \Big) \Big\}
\end{equation}

\begin{equation}\label{eq4}{\lambda_l^2}^{t+1} = \max \Big\{0, {\lambda_l^2}^t + a_t \Big( -\overline{d_l} + \sum\limits_{k=1}^m q_{kl}^t \Big) \Big\}
\end{equation}

\noindent where the sequence $\{a_t\}$ must satisfy $a_t > 0, a_t \rightarrow 0$ y $\sum\limits_{t=0}^{\infty} a_t = \infty$ to guarantee the convergence of the iterative scheme \cite{dupuis1993dynamical}.

When designing the P system that will model these equations, the main problem to solve is dealing with the two partial derivatives in Equation \ref{eq1}. The discrete encoding of the information of P systems as multisets of objects is an added difficulty in order to deal with derivatives, therefore, we have made the following decisions in the design of the P system:
\begin{itemize}
\item The functions $P_j(q)$, which represents the funds in donation dollars due to the visibility of all NGOs at location $j$, in many cases can be expressed as 

$$P_j(q) = k_j \sqrt{\sum\limits_{i=1}^m q_{ij}}$$ 

\noindent with $k_j > 0$, thus $$\frac{\partial P_j(q^t)}{\partial q_{kl}} = 0 \mbox{ if $j \neq l$}$$ 

\noindent and 
$$\frac{\partial P_j(q^t)}{\partial q_{kl}} = \frac{k_j}{2} (\sum\limits_{i=1}^m q_{ij})^{-1/2} \mbox{ when $j = l$}$$ 

This usually gives the term $$\frac{\partial P_j(q^t)}{\partial q_{kl}}$$ 

\noindent various orders of magnitude lower than the other terms on Equation \ref{eq1}. Because of this, we have decided to ignore it. In fact, in Section \ref{section:experiments}, we experimentally show that the impact derived from this decision is low.

\item Similarly, the cost functions are of the form $$c_{ij}(q_{ij}) = (a_{kl} q_{ij} + b_{kl})^2$$ 

\noindent with $a_{kl}, b_{kl} > 0$, so $\frac{\partial c_{kl}(q_{kl}^t)}{\partial q_{kl}} = 2 a_{kl}^2 q_{kl} + 2 a_{kl} b_{kl}$. Finally, we can rewrite then Equation \ref{eq1} as: 

\begin{equation}\label{eq1bis}
    q_{kl}^{t+1} = \max \Big\{0, \big\{ q_{kl}^t + a_t \Big( \frac{\omega_k \gamma_{kl}}{\beta_k} - \frac{1}{\beta_k} (2 a^2_{kl} q_{kl} + 2 a_{kl} b_{kl}) - \lambda_k^t + {\lambda_l^1}^t - {\lambda_l^2}^t \Big) \big\} \Big\}
\end{equation}
\end{itemize}

We note that the conditions that guarantee the existence and uniqueness of the solution have not been compromised. In Section \ref{section:design} a P system that simulates the algorithm defined by Equations \ref{eq2}, \ref{eq3}, \ref{eq4} and \ref{eq1bis} is designed.





\section{Transition P Systems with Membrane Polarization}\label{Transition}

In this section, we introduce the chosen P system model for designing our solution to the GNE problems. We have chosen the model of transition P systems \cite{DBLP:journals/jcss/Paun00} endowed with membrane polarization, i.e.,
the membranes have a charge that acts as a state of the membrane. That state controls the evolution of the objects within the membrane. For some definitions about membrane systems and formal languages, we refer the reader to~\cite{HandbookMC10,rozenberg1997handbook}. A transition P system with membrane polarization of degree $q \geq 1$ is a tuple $$\Pi = (\Gamma, \mu, \mathcal{M}_1, \ldots, \mathcal{M}_q, (\mathcal{R}_1, \rho_1), \ldots, (\mathcal{R}_q, \rho_q), i_{out}),$$ where:

\begin{enumerate}
    \item $\Gamma$ is an alphabet whose elements are called objects;
    \item $\mu$ is a hierarchical tree-like structure;
    \item $\mathcal{M}_1, \ldots, \mathcal{M}_q$ are multisets of objects over $\Gamma$;
    \item $\mathcal{R}_1, \ldots, \mathcal{R}_q$ are sets of rules of the following form:
    \begin{itemize}
        \item $[ \, u \rightarrow v \, ]^\alpha_h, u, v \in M(\Gamma), h \in \{1, \ldots, q\}, \alpha \in \{0, +, -\}$ is an object evolution rule;
        \item $[ \, u \, ]^\alpha_h \rightarrow v \, [\quad]^\beta_h, u, v \in M(\Gamma), h \in \{1, \ldots, q\}, \alpha, \beta \in \{0, +, -\}$ is an send-out communication rule;
        \item $u \, [\quad]^\alpha_h \rightarrow  [ \, v \, ]^\beta_h, u, v \in M(\Gamma), h \in \{1, \ldots, q\}, h$ is not the label of the skin membrane$, \alpha, \beta \in \{0, +, -\}$ is an send-in communication rule;
    \end{itemize}
    \item $\rho_1, \ldots, \rho_q$ are partial weak relations between rules from $\mathcal{R}_1, \ldots, \mathcal{R}_q$, respectively;
    \item $i_{out} \in \{0, 1, \ldots, q\}$ is the output region (if $i_{out} = 0$, the output region is the \emph{environment} of the system).
\end{enumerate}

A transition P system with membrane polarization $$\Pi = (\Gamma, \mu, \mathcal{M}_1, \ldots, \mathcal{M}_q, (\mathcal{R}_1, \rho_1), \ldots, (\mathcal{R}_q, \rho_q), i_{out})$$ can be seen as a set of $q$ membranes organized in a rooted tree-like structure whose root node is called the skin membrane, each of them having a polarization among $0$, $+$, or $-$. A configuration of $\Pi$ in an instant $t$ can be described by the multisets of objects in each membrane in such a moment and the polarization of each membrane and the multiset of objects in the environment, denoted by $\mathcal{M}_{0,t}$; that is, $C_t = ((\mathcal{M}_{1,t}, \alpha_{1,t}), \ldots, (\mathcal{M}_{q,t}, \alpha_{q,t}), \mathcal{M}_{0,t})$. It can also be described in a more graphical way such that $[u]^\alpha_h$ represents that membrane $h$ contains the multiset of objects $u$ and its charge is $\alpha$. If the graphical description contains a membrane $h'$ such that it is in the membrane $h$, like $[ \, [\quad]_{h'} \, ]_h$, then $h'$ is a \emph{child} membrane of $h$. The initial configuration of $\Pi$ is $C_0 = ((\mathcal{M}_1, 0), \ldots, (\mathcal{M}_{q}, 0), \emptyset)$. A configuration is called a \emph{halting configuration} if no more rules are applicable to it. For the sake of simplicity, we will use the notation $\mathbb{C}_t$ to denote specific parts of the P system in such a way that $\mathbb{C}_t = \mu'$, where $\mu'$ is a subtree of $\mu$, and the multisets of objects associated to each membrane $h$ of $\mu'$ at an instant $t$ is denoted by $[ \, u \, ]_h$, where $u$ is a string that represents the multiset $\mathcal{M}_{h,t}$.

An object evolution rule $[ \, u \rightarrow v \, ]^\alpha_h, u, v \in M(\Gamma), h \in \{1, \ldots, q\}, \alpha \in \{0, +, -\}$ is applicable to a configuration $C_t$ if there exists a membrane labeled by $h$ in $\Pi$ such that it contains the multiset of objects $u$ and its polarization is $\alpha$. Applying such a rule leads to the removal of $u$ from the membrane $h$ and the generation of the multiset of objects $v$ in the membrane $h$. 

An send-out communication rule $[ \, u \, ]^\alpha_h \rightarrow v \, [\quad]^\beta_h, u, v \in M(\Gamma), h \in \{1, \ldots, q\}, \alpha, \beta \in \{0, +, -\}$ is applicable to a configuration $C_t$ if there exists a membrane labelled by $h$ in $\Pi$ such that it contains the multiset of objects $u$ and its polarization is $\alpha$. The application of such a rule leads to the removal of $u$ from such a membrane $h$, the generation of the multiset of objects $v$ in the parent region of $h$ and the change of polarization of such a membrane $h$ from $\alpha$ to $\beta$.

An send-in communication rule $u \, [\quad]^\alpha_h \rightarrow  [ \, v \, ]^\beta_h, u, v \in M(\Gamma), h \in \{1, \ldots, q\}, h$ is not the label of the skin membrane$, \alpha, \beta \in \{0, +, -\}$ is applicable to a configuration $C_t$ if there exists a membrane labelled by $h$ in $\Pi$ such that its parent membrane contains the multiset of objects $u$ and the polarization of such membrane $h$ is $\alpha$. The application of such a rule leads to the removal of $u$ from the parent membrane, the generation of the multiset of objects $v$ in such a membrane $h$, and the change of polarization of such a membrane $h$ from $\alpha$ to $\beta$.

In addition, if $(r_1, r_2) \in \rho_h$, it means that $r_1$ has priority over $r_2$ in the following sense: Rule $r_2$ is applicable only if the remaining objects from membrane $h$ cannot fire rule $r_1$ any more times. If one object can fire more than one rule, then it will be selected in a non-deterministic way.

Apart from that, object evolution rules are applied in a maximal parallel way; that is, any object that can fire an applicable rule will fire it. A multiset of rules $U$ is maximal if there is no other multiset of rules $U'$ in the set of multisets of applicable rules such that $U \subset U'$.

Send-in and send-out rules are applied in a maximal parallel way as follows: Let $r_1$ and $r_2$ two rules where either $r_1$ and $r_2$ can be a send-in or a send-out rule. Let $h_1$ ($h_2$, respectively) be the label of the membrane affected by the rule $r_1$ ($r_2$, resp.). Let $\alpha_1$ ($\alpha_2$, respectively) be the polarization of the left-hand side of the rule (that is, the part of the rule that is to the left of the arrow) of rule $r_1$ ($r_2$, resp.), and let $\beta_1$ ($\beta_2$, resp.) be the polarization of the right-hand side of the rule $r_1$ ($r_2$, resp.). We say that $r_1$ and $r_2$ are compatible if the following holds: $h_1 = h_2 \land \alpha_1 = \alpha_2 \rightarrow \beta_1 = \beta_2$; that is, two rules that affect the same membrane and can be applied in the same moment $t$ are compatible if and only if the resulting polarization by the application of each rule is the same. Each multiset of applicable rules can contain only compatible rules.

A \emph{transition} or a \emph{computational step} of $\Pi$ is made by applying the rules in the aforementioned way in all membranes at the same time, and it is denoted by $C_t \Rightarrow_\Pi C_{t+1}$.

A \emph{computation} of $\Pi$ is a sequence of configurations $\mathcal{C} = (C_0, C_1, \ldots, C_n)$ such that $C_0$ is the initial configuration of $\Pi$ and for each $t$, $C_t \Rightarrow C_{t+1}$. A computation is \emph{halting} if $n \in \mathbb{N}$, and $C_n$ is a halting configuration.

\subsection{An example of a P system}

For the sake of simplicity, we give a simple example to explain the behavior of transition P systems with membrane polarization. Let $$\Pi = (\Gamma, \mu, \mathcal{M}_1, (\mathcal{R}_1, \rho_1), i_{out})$$ be a transition P system with membrane polarization of degree $1$ where:
\begin{enumerate}
    \item $\Gamma = \{a, b, c, d, e\}$;
    \item $\mu = [\quad]_1$;
    \item $\mathcal{M}_1 = \{a^3, d\}$
    \item $\mathcal{R}_1 = \{r_1 \equiv [ \, a^2 \rightarrow b \, ]_1, r_2 \equiv [ \, a \rightarrow c \, ]_1, r_3 \equiv [ \, d \rightarrow e \, ]_1\}$;
    \item $\rho_1 = \{(r_1, r_2)\}$;
    \item $i_{out} = 0$.
\end{enumerate}

The initial configuration can be represented as $\mathbb{C}_0 = [ \, a^3, d \, ]^0_1$. Then, the following multisets of rules are applicable:
\begin{itemize}
    \item $U_1 = \{r_1, r_2\}$;
    \item $U_2 = \{r_1\}$;
    \item $U_3 = \{r_1, r_3\}$;
    \item $U_4 = \{r_1, r_2, r_3\}$;
\end{itemize}
Since we look for maximal multisets of applicable rules, we can observe that $U_1, U_2, U_3 \subset U_4$. Therefore, the multiset $U_4$ will be applied in the first computational step, leading to the following configuration: $\mathbb{C}_1 = [ \, b, c, e \, ]_1$.

\section{Design and functioning of the P system}\label{section:design}

Let us consider the Equations \ref{eq2}, \ref{eq3}, \ref{eq4} and \ref{eq1bis} defined in Section \ref{sectioneuler}. In this section, a P system that computes the solutions of these equations is analyzed\footnote{A detailed description of the P system can be found in the Appendix \ref{appendix:DefinitionPSystem}.}. 

We first need to define a sequence $\{a_t\}$ that satisfies 
$$a_t > 0, a_t \rightarrow 0 \mbox{ and } \sum\limits_{t=0}^{\infty} a_t = \infty$$ 

\noindent as indicated in Section \ref{sectioneuler}. The sequence $a_t = 0.1 \cdot b_t$ is defined by {\small$\{b_t\} =$  $$\{ \underbrace{1,1, \ldots, 1}_{1024 \text{ times}},           \underbrace{1/2,1/2, \ldots, 1/2}_{1024 \text{ times}}, \ldots , \underbrace{1/1024,1/1024, \ldots, 1/1024}_{1024 \text{ times}}, \underbrace{1/2048,1/2048, \ldots, 1/2048}_{2048 \text{ times}}, \ldots \}$$} taking $r$ copies of the element $0.1 \cdot 1/2^w$ with $r = \max\{1024, 2^w\}$ and $w = 0,1,\dots$. This sequence guarantees convergence and can be easily updated and employed using the evolution rules of a P system. In opposition, the implementation of the sequence used in \cite{NagurneyMain} can raise the complexity of the model. The terms $a_t$ are represented through objects $s$ in the P system.

The computation of the P system can be summarized in a loop of three stages, represented in Algorithm \ref{alg:psystem}.

\begin{algorithm}
\caption{General overview of the P system computation}\label{alg:psystem}
\begin{algorithmic}
\While{True}
\State 1. \textbf{Initialization}.
\State 2. \textbf{Update} of values $q_{kl}^{t+1}, \lambda_k^{t+1}, \lambda_l^{1^{t+1}}$ and $\lambda_l^{2^{t+1}}$. If necessary, update $a_t$.
\State 3. \textbf{Comparison} of values $q_{kl}^{t+1}$ with values $q_{kl}^{t}$. If no difference is found, STOP.
\EndWhile
\end{algorithmic}
\end{algorithm}

Next, an analysis of the computation of each stage is provided. The result of the analysis is that, for each time step, the number of transition steps in the Initialization and Comparison stages is bounded by a constant. Meanwhile, the number of transition steps in the Update stage is constant for the first 10240 iterations (1024 identical elements for each of the first 10 values in the sequence $a_t$) and then grows logarithmically with time. This means that the time complexity of the global computation only depends on the number of iterations required by the original algorithm.

In the computation analysis, $\mathbb{C}_\tau$ represents the $\tau$-th configuration of each iteration; that is, $\mathbb{C}_0$ is the initial configuration in each iteration. The main difference between iterations is given by the existence of objects $s$, and this fact is considered along the computation. \newline

\textbf{Stage 1: Initialization } \\ In this stage, all the necessary objects for updating the values $q_{kl}^{t+1}, \lambda_k^{t+1}, \lambda_l^{1^{t+1}}$ and $\lambda_l^{2^{t+1}}$ are created and distributed to the corresponding membranes. We assume that this stage starts with some objects $x_{k,l}, la_k, la_{1,l}, la_{2,l}$ inside the membrane $INIT$ which represent the values $q_{kl}^{t}, \lambda_k^{t}, \lambda_l^{1^{t}}$ and $\lambda_l^{2^{t}}$ at a time step $t$. For simplicity, we consider only the objects $x_{k,l}$, but the behavior of the other objects is analogous. If we only consider the membranes involved in each step, we have the following configuration:
    $$ \mathbb{C}_0 = [[x_{k,l}^{q_{k,l}}]_{INIT}^0 \ y_0 \ ]_1^0 $$
In the next transition step, different copies of each object in the membrane $INIT$ are ejected into the skin membrane. The objects $y$ are applied to coordinate the computation:
    $$ \mathbb{C}_1 = [x_{k,l}^{q_{k,l}}, xt_{k,l}^{q_{k,l}}, xl_{0,k,l}^{q_{k,l}}, xl_{1,k,l}^{q_{k,l}}, xl_{2,k,l}^{q_{k,l}}, y_{0,k,l}, yl_{k}, yl_{1,l}, yl_{2,l} ]_1^0 $$
In one transition step, objects representing the constant part of \Cref{eq2,eq3,eq4,eq1bis} (before multiplying by $a_t$)  are created in the corresponding membranes. For simplicity, only the membranes $Q_{k,l}$, which represent the Equation \ref{eq1bis}, are considered. The objects $p_0$ represent originally the term $\frac{\omega_k \gamma_{kl}}{\beta_k}$, while objects $ct_0$ represent the term $\frac{-2 a_{kl} b_{kl}}{\beta_k}$ (see rule $RS_{1,6}$ in the 
Appendix \ref{appendix:DefinitionPSystem}).  In one transition step:
    $$ \mathbb{C}_2 = [ \ [ y_{0}, p_0^{\kappa_0}, ct_0^{\kappa_1} ]^-_{Q_{k,l}} \ x_{k,l}^{q_{k,l}}, xt_{k,l}^{q_{k,l}}, xl_{0,k,l}^{q_{k,l}}, xl_{1,k,l}^{q_{k,l}}, xl_{2,k,l}^{q_{k,l}}  ]_1^0 $$
Now that the charge of the membranes is negative, objects representing $q_{kl}^{t}, \lambda_k^{t}, \lambda_l^{1^{t}}$ and $\lambda_l^{2^{t}}$ are inserted into the corresponding membranes. The terms $q_{kl}^t$ are represented by the objects $p$ and $c_0$, while the $\lambda$-terms are represented by objects $p_0$ or $n_0$. These objects will be used to update the variables using \Cref{eq2,eq3,eq4,eq1bis} (see rules $RS_{1,10}$ to $RS_{1,19}$ in the Appendix \ref{appendix:DefinitionPSystem}). In one transition step:
    $$ \mathbb{C}_3 = [ \ [ y_{1}, p_0^{\kappa_0}, ct_1^{\kappa_1},  p^{q_{k,l}}, c_0^{q_{k,l}} ]^-_{Q_{k,l}} ]_1^0 $$
In this step, all membranes $Q_{k,l}$, $LAMB_k$, $LAMB_{1,l}$ and $LAMB_{2,l}$ have the necessary objects to start the Update stage, and all of them are coordinated through the usage of the object $y_1$. The Initialization stage is then completed in three transition steps. \newline

\textbf{Stage 2: Update } \\ In this stage, the values $q_{kl}^{t+1}, \lambda_k^{t+1}, \lambda_l^{1^{t+1}}$ and $\lambda_l^{2^{t+1}}$ are computed following \Cref{eq2,eq3,eq4,eq1bis}. The computation has two phases: one of computing the parentheses in each equation and multiplying it by $a_t$, and one of computing the $\max$ function. For simplicity, only membranes $Q_{k,l}$ are considered, but the behavior for membranes $LAMB_k$, $LAMB_{1,l}$ and $LAMB_{2,l}$ is analogous. This stage starts with the configuration $\mathbb{C}_3$. In two transition steps, all terms in the parentheses of \Cref{eq2,eq3,eq4,eq1bis} will be represented by objects $p_0$ or $n_0$ depending on their sign:
$$ \mathbb{C}_5 = [ \ [ y_{3}, p_0^{\kappa_0}, n_0^{\kappa_1 + \lfloor 2 \cdot a^2_{kl}  / \beta_k \rfloor \cdot q_{k,l}},  p^{q_{k,l}} ]^-_{Q_{k,l}} ]_1^0 $$

In one transition step, objects $y_3$ will be introduced in the membranes $REDUCE$, changing their charge to negative. In one more transition step, objects $p_0$ and $n_0$ will be introduced in the membranes $REDUCE$. For the first 1024 iterations, in the next step objects $p_0$ and $n_0$ will be added, generating objects $p$ or $n$, and the object $y_6$ will be generated, changing the charge of membranes $REDUCE$ to positive. In the rest, there will be objects $s$ present that will cause the reduction in the number of objects $p$ and $n$ by multiplying them by $10 \cdot a_t$. There is one more transition step for each object $s$ present, and there are $n$ objects $s$ present for the sequence term $a_t = 0.1 \cdot 1/2^w$. Assuming we end with objects $p$ (analogous otherwise), in at least three transition steps we have the configuration:
$$ \mathbb{C}_{\geq 8} = [ \ [ y_{6} \ [p^{10 \cdot a_t \cdot (\kappa_0 - \kappa_1 - \lfloor 2 \cdot a^2_{kl}  / \beta_k \rfloor \cdot q_{k,l})}]^+_{REDUCE_{0,k,l}}  p^{q_{k,l}} ]^-_{Q_{k,l}} ]_1^0 $$
In two more transition steps, the objects $y_6$ will evolve and change the charge of membranes $REDUCE$ back to neutral and the charge of membranes $Q_{k,l}, LAMB_k$, $LAMB_{1,l}$ and $LAMB_{2,l}$ to positive. During these steps, the number of objects $p$ or $n$ that were previously in membranes $REDUCE$ will be multiplied by $0.1$ and added to the rest in membranes $Q_{k,l}, LAMB_k$, $LAMB_{1,l}$ and $LAMB_{2,l}$, updating the terms $q_{kl}^{t+1}, \lambda_k^{t+1}, \lambda_l^{1^{t+1}}$ and $\lambda_l^{2^{t+1}}$:
$$ \mathbb{C}_{\geq 10} = [ \ y_{8,k,l} \ [ \ [ \ ]^0_{REDUCE_{0,k,l}} p^{q_{kl}^{t+1}} ]^+_{Q_{k,l}} ]_1^0 $$
In the two next transition steps, if objects $p$ and $n$ are present they cancel each other. If objects $p$ are left, they are ejected to the skin as objects $o_{kl}$, $i_{kl}$ and $xt_{1,k,l}$, and if objects $n$ are left they are deleted. 
$$ \mathbb{C}_{\geq 12} = [ \ y_{10}^{m \cdot n},  o_{k,l}^{(q_{k,l}^{t+1})}, i_{k,l}^{(q_{kl}^{t+1})}, xt_{1,k,l}^{(q_{kl}^{t+1})} \ [ \ [ \ ]^0_{REDUCE_{0,k,l}}  ]^0_{Q_{k,l}} ]_1^0 $$


The Update stage is then completed in a minimum of nine transition steps, and the number of transition steps grows linearly to nineteen for the 10240 first iterations and logarithmically with $t$ for the rest. In practice, a solution with convergence tolerance $10^{-10}$ is found in the first 2048 iterations, so we have an effective upper bound of ten transition steps. \newline

\textbf{Stage 3: Comparison } \\ In this stage, the new values $q_{kl}^{t+1}$ are compared with the previous values $q_{kl}^{t}$. If a difference is found, the execution continues. Otherwise, the new values are sent to the membrane $OUTPUT$, and the computation finishes. This stage starts with the following configuration:
$$ \mathbb{C}_{\geq 12} = [ \ y_{10}^{m \cdot n},  o_{k,l}^{(q_{k,l}^{t+1})}, i_{k,l}^{(q_{kl}^{t+1})}, xt_{1,k,l}^{(q_{kl}^{t+1})}, xt_{k,l}^{(q_{kl}^{t})} \ [ \ ]^0_{COMP} \ [ \ ]^0_{INIT} \ [ \ ]^0_{OUTPUT} 
 \ ]_1^0 $$
 In two transition steps, objects $xt_{k,l}$ and $xt_{1,k,l}$ are in membrane $COMP$ to be compared: 
 $$ \mathbb{C}_{\geq 14} = [ \  i_{k,l}^{(q_{kl}^{t+1})} \ [ \ y_{12}, o_{1,k,l}^{(q_{k,l}^{t+1})}, xt_{1,k,l}^{(q_{kl}^{t+1})}, xt_{k,l}^{(q_{kl}^{t})} \ ]^0_{COMP} \ [ \ ]^0_{INIT} \ [ \ ]^0_{OUTPUT} 
 \ ]_1^0 $$
 In the next two transition steps, objects $xt_{k,l}$ and $xt_{1,k,l}$ cancel each other. If at least one of them is present, then $y_{16}$ evolves to $y_{13}$ and is ejected as $y_{14}$. Otherwise, an object $stop$ is created and ejected. The rest of the objects $xt_{k,l}$ and $xt_{1,k,l}$ are deleted.
 $$ \mathbb{C}_{\geq 16} = [ \  o_{3,k,l}^{(q_{k,l}^{t+1})}, i_{k,l}^{(q_{kl}^{t+1})}, y_{14}||stop \ [ \  ]^0_{COMP} \ [ \ ]^0_{INIT} \ [ \ ]^0_{OUTPUT} 
 \ ]_1^0 $$
 At this moment, two things can happen. The first one is that in the next three transition steps, the $stop$ object changes the membrane $OUTPUT$ charge to negative, the objects $o_{3,k,l}$ go into it, and the computation finishes after changing again the charge of $COMP$ to neutral:
 $$ \mathbb{C}_{\geq 19} = [ \ i_{k,l}^{(q_{kl}^{t+1})} \ [ \ ]^0_{COMP} \ [ \ ]^0_{INIT} \ [ o_{k,l}^{(q_{k,l}^{t+1})} ]^0_{OUTPUT} 
 \ ]_1^0 $$
 The other possibility is that, in the following two transition steps, the $y_{14}$ object changes the charge of membrane $INIT$, the objects $i_{k,l}$ get into membrane $INIT$, and an object $y_{0}$ is created, leaving the P system in a state where a new iteration can start again:
 $$ \mathbb{C}_{\geq 18} = [  y_0 [ \ ]^0_{COMP} \ [ x_{k,l}^{(q_{kl}^{t+1})}, count_0 \ ]^0_{INIT} \ [ \ ]^0_{OUTPUT} 
 \ ]_1^0 $$
 In this last step, a $count_0$ object is also created. This object will contribute to the update of $a_t$.
 We can see then that this stage is completed in six transition steps except for the last iteration, which is completed in seven steps.

 The result of this analysis is that the whole computation of the system only depends on the number of iterations required by the original algorithm. The algorithmic time complexity of an iteration of the algorithm is constant for Stage 1, logarithmic with $t$ for Stage 2, and constant for Stage 3. The algorithmic time complexity of the whole computation is then $\mathcal{O}(t \ log(t))$. In practice, it is $\mathcal{O}(t)$, as explained in Stage 2. In contrast, the original algorithm has to compute Equations \ref{eq2}, \ref{eq3}, \ref{eq4}, and \ref{eq1bis} at every iteration, so the algorithmic time complexity of the original algorithm is $\mathcal{O}((m \cdot n + m + 2n) \cdot t)$. Consequently, we present a reduction of a factor of $m \cdot n + m + 2n$.

\section{Experiments}\label{section:experiments}

Different P system simulators are available, such as P-lingua (MeCoSim) \cite{DBLP:journals/nc/Gutierrez-NaranjoPR08, MecosimRef16, MecosimRef17} or UPSimulator \cite{UPSimulator, UPSimulator2}. To perform experiments, MeCoSim has been updated and chosen to implement our P system. The last simulator update allows for the design of transition P systems with membrane polarization. 

To test the correct behavior of our P system, 
we have considered several examples taken from \cite{NagurneyMain}, where the numerical data of some case studies are provided. Namely, the authors provide four toy examples and the real-world case study of Hurricane Katrina. All of these examples have been simulated. For the five numerical examples, the convergence tolerance chosen is $10^{-5}$, while for the Hurricane Katrina case study, the tolerance is $10^{-10}$. A solution was found for all the numerical examples in less than 1024 iterations. The Update stage took then nine transition steps for every iteration. For the Hurricane Katrina case study, a solution was found in less than 1100 iterations. The Update stage took then nine transition steps for the first 1024 iterations, and ten transition steps for the rest (see Section \ref{section:design}).

The results of the simulations of the four toy numerical examples can be found in Tables \ref{Tab:ej1}, \ref{Tab:ej2}, \ref{Tab:ej3} and \ref{Tab:ej4}, where the obtained values for $q_{ij}$ (i.e., the number of items that each NGO $i$ provides to a demand point $j$) are compared. On the right, the equilibrium values obtained in \cite{NagurneyMain} are shown, and, on the left, the values obtained with our P system. Considering that the true values are rounded to one decimal, it can be observed that the results returned by our P system coincide perfectly with the true values, even after replacing Equation \ref{eq1} with Equation \ref{eq1bis} as discussed in Section \ref{sectioneuler}. 

Regarding the Hurricane Katrina case study, the replacement of Equation \ref{eq1} with Equation \ref{eq1bis} has impacted the solutions returned, as can be observed in Table \ref{Tab:Katrina}. However, the solution returned by the P system is a good approximation: the average error between the correct solution and the solution returned by our P system is 1.98\%, the median is 0.82\%, the maximum error is 7.89\%, and only in four cases out of thirty is the error higher than 5\%. 

These experiments support our claim that the contribution of the functions $P_j(q)$ can be ignored without incurring significant errors. This implies that the financial funds due to the visibility of all NGOs at each location have little impact on the final solution of the problem of how to distribute humanitarian relief subject to the constraints stated if they follow the assumptions in \cite{NagurneyMain}.

\begin{table}
\captionsetup{width=.75\textwidth}
\caption{Results of the simulation for Example 1 of \cite{NagurneyMain}.}
\centering
\begin{tabular}{ c c c }
    \hline
    $q_{ij}$ & P system & Solution \\
    \hline
    $q_{11}$ & 352.50012 & 352.5 \\
    $q_{21}$ & 247.50012 & 247.5 \\
    \hline
\end{tabular}
\label{Tab:ej1}
\caption{Results of the simulation for Example 2 of \cite{NagurneyMain}.}
\centering
\begin{tabular}{ c c c }
    \hline
    $q_{ij}$ & P system  & Solution \\
    \hline
    $q_{11}$ & 352.50012 & 352.5 \\
    $q_{12}$ & 452.50004 & 452.5 \\
    $q_{21}$ & 247.50012 & 247.5 \\
    $q_{22}$ & 347.50004 & 347.5 \\
    \hline
\end{tabular}
\label{Tab:ej2}

\caption{Results of the simulation for Example 3 of \cite{NagurneyMain}.}
\centering
\begin{tabular}{c c c}
    \hline
    $q_{ij}$ & P system  & Solution \\
    \hline
    $q_{11}$ & 423.75003 & 423.8 \\
    $q_{12}$ & 471.25002 & 471.3 \\
    $q_{13}$ & 436.87498 & 436.9 \\
    $q_{21}$ & 176.25011 & 176.3 \\
    $q_{22}$ & 328.75007 & 328.8 \\
    $q_{23}$ & 563.12492 & 563.1 \\
    \hline
\end{tabular}
\label{Tab:ej3}

\caption{Results of the simulation for Example 4 of \cite{NagurneyMain}.}
\centering
\begin{tabular}{c c c}
    \hline
    $q_{ij}$ & P system  & Solution \\
    \hline
    $q_{11}$ & 411.25004 & 411.3 \\
    $q_{12}$ & 458.75002 & 458.8 \\
    $q_{13}$ & 499.37496 & 499.4 \\
    $q_{21}$ & 138.75012 & 138.8 \\
    $q_{22}$ & 291.25007 & 291.3 \\
    $q_{23}$ & 750.62488 & 750.6 \\
    \hline
\end{tabular}
\label{Tab:ej4}
\end{table}

\begin{sidewaystable}

\caption{Results of the simulation for the use case of Hurricane Katrina in \cite{NagurneyMain}.}
\label{Tab:Katrina}

\begin{tabular}{c c c c @{\hspace{1cm}} c c c @{\hspace{1cm}} c c c}
    \hline
    &  \multicolumn{3}{c}{Others} & \multicolumn{3}{c}{Red cross} & \multicolumn{3}{c}{Salvation army} \\
    \cmidrule(lr){2-4}\cmidrule(lr){5-7}\cmidrule(lr){8-10}
    Location & P system & Solution & Error &  P system & Solution & Error &  P system & Solution & Error  \\
    \hline
    St.Charles & 16.10 & 17.48 & \textbf{7.89\%} & 29.47 & 28.89 & 2.01\% & 4.35 & 4.192 & 3.77\% \\
    Terrebonne & 267.93 & 267.02 & 0.34\% & 410.31 & 411.67 & 0.33\% & 73.38 & 73.57 & 0.26\% \\
    Assumption & 47.92 & 49.02 & 2.24\% & 77.59 & 77.26 & 0.43\% & 13.09 & 12.97 & 0.93\% \\
    Jefferson & 264.57 & 263.69 & 0.33\% & 405.34 & 406.68 & 0.33\% & 72.30 & 72.45 & 0.21\% \\
    Lafourche & 186.55 & 186.39 & 0.09\% & 287.22 & 287.96 & 0.26\% & 51.11 & 51.18 & 0.14\% \\
    Orleans & 466.04 & 463.33 & 0.58\% & 710.66 & 713.56 & 0.41\% & 126.63 & 127.1 & 0.37\% \\
    Plaquemines & 20.53 & 21.89 & \textbf{6.21\%} & 37.02 & 36.54 & 1.31\% & 4.37 & 4.23 & 3.31\% \\
    St.Barnard & 74.52 & 72.31 & 3.06\% & 120.12 & 115.39 & 4.10\% & 17.13 & 16.22 & \textbf{5.61\%} \\
    St.James & 57.66 & 58.67 & 1.72\% & 92.31 & 92.06 & 0.27\% & 15.77 & 15.66 & 0.70\% \\
    St.John Baptist & 16.83 & 18.2 & \textbf{7.52\%} & 30.60 & 29.99  & 2.03\% & 4.51 & 4.40 & 2.50\% \\
    \hline
\end{tabular}


\end{sidewaystable}

\section{Technical Conclusions}
The simulation of continuous processes by intrinsically discrete models is always a complex task, and each real-world problem needs to be deeply studied in order to adopt the best possible solution. In this paper, we have considered the minimization question expressed in Eq. \ref{eq1}. Such an equation involves two terms with derivatives. After a deep study of the problem, we can conclude that the term with the first derivative is several orders of magnitude lower than the remaining terms. By removing this term, we lose accuracy; however, from a practical point of view, this loss is not significant, as the experiments show. Concerning the second term with a derivative, a solution for a general function $c_{ij}$ is not possible, but it can be reached in this case bearing in mind that it can be expressed as a second-grade polynomial equation.

From a Membrane Computing perspective, we would like to emphasize that our design shows that the number of transition steps in the Initialization and Comparison stages is bounded by a constant and the number of transition steps in the Update stage grows linearly for the first 10240 iterations and then grows logarithmically with time. This can be considered as the main contribution of this paper from the theoretical side. Due to the intrinsic massive parallelism of the Membrane Computing devices, the time complexity of the global computation, considered on the basis of the P system steps, only depends now on the number of iterations required by the original algorithm. Besides, the modularity of the design allows us to extend the process with new agents by introducing new modules and only minimal changes in the other modules. Additionally, the object-based approach is similar to agent-based models in the sense that the behaviors of these individuals can be tracked, having a one-to-one relationship between objects and real-life resources, and leading to a fine-grained model that can be calibrated for different scenarios.

As a final remark, it is worth stressing that, although the use of transition P systems with polarizations to Generalized Nash Equilibria was introduced in 
\cite{LuqueCerpa2024}, a first detailed real-life use case, specifically applied to model humanitarian relief, is presented here.

\section{Final remarks}
In this paper, we have shown that Membrane Computing is a useful computational paradigm to model humanitarian relief. The main contribution of this paper is twofold. On the one hand, we contribute to highlighting that it is unacceptable that efforts to distribute humanitarian aid are not enough successful due to a lack of coordination,
congestion in terms of logistics, and duplication of services. This situation demands an answer from the scientific community, and many theoretical efforts must be made to optimize the resources. One possible solution is to model the problem in terms of GNE, but other optimization methods can be explored. On the other hand, to the best of our knowledge, this is the first paper bridging Membrane Computing with the distribution of humanitarian relief and we have shown that P systems can be a useful tool to model complex situations in this area.

As future research, many applications of Membrane Computing to logistics in humanitarian aid can be explored, not only with transition P systems but with other P systems models. In parallel, other optimization methods can be considered to optimize humanitarian aid distribution. As a consequence of the good behavior of the model, it seems reasonable to study specific situations where the general case is not enough to model real-life events, such as abnormal donation patterns, changes in the political landscape, or any other fact that could impact the distribution of humanitarian relief.

\backmatter

\bmhead{Supplementary information}

Not applicable

\bmhead{Acknowledgements}
M.A.G.N. acknowledges the support of the project {\it REliable \& eXplAinable Swarm Intelligence for People with Reduced mObility (REXASI-PRO)} founded by the European Commission $-$ GRANT AGREEMENT NO. 101070028










\begin{appendices}

\section{Appendix: Definition of the P system \label{appendix:DefinitionPSystem}}

Let $\mathcal{N} = \{1,...,m\}$ and $\mathcal{D} = \{1,...,n\}$ with $m$ and $n$ the number of NGOs and demand locations respectively. Let $P^{-1} = 10^{-p}$ with $p \in \mathbb{N}$ be the tolerance for convergence of the algorithm. Let  us consider the following transition P system with membrane polarization $$ \Pi = \langle \Gamma, H, EC, \mu, \{w_h\}_{h \in H}, (\mathcal{R},\rho) \rangle $$
where the alphabet of objects is given by:
\begin{align*}
    \Gamma &= \{ y_0, y_1, y_2, y_3, y_4, y_5, y_6, y_7, y_{10}, y_{11}, y_{12}, y_{13}, y_{14}, y_{15} \} \cup \\
    &\cup \{ y_{0,k,l}, yl_{0,k}, yl_{1,l}, yl_{2,l} \ | \ k \in \mathcal{N}, l \in \mathcal{D} \} \  \cup \\
    &\cup \{y_{8,k,l}, y_{9,k,l}, yla_{8,k}, yla_{9,k}, yla_{1,8,l}, yla_{1,9,l}, yla_{2,8,l}, yla_{2,9,l} \ | \ k \in \mathcal{N}, l \in \mathcal{D} \} \  \cup \\
    &\cup \{x_{k,l}, xt_{k,l}, xt_{1,k,l}, xl_{0,k,l}, xl_{1,k,l}, xl_{2,k,l} \ | \ k \in \mathcal{N}, l \in \mathcal{D} \} \  \cup \\
    &\cup \{la_{k}, laq_{0,k,l}, la_{0,k}, la_{1,l}, laq_{1,k,l}, la_{1,l}, la_{2,l}, laq_{2,k,l}, la_{2,l} \ | \ k \in \mathcal{N}, l \in \mathcal{D} \} \  \cup \\
    &\cup \{ p_0, n_0, ct_0, ct_1, ct_2, c_0, c_1, p, n, o \} \  \cup \\
    &\cup \{i_{k,l}, lao_{0,k}, lao_{1,l}, lao_{2,l}, o_{k,l}, o_{0,k,l}, o_{1,k,l}, o_{2,k,l}, o_{3,k,l}, o_{4,k,l}, \ | \ k \in \mathcal{N}, l \in \mathcal{D} \} \  \cup \\
    &\cup \{ rem, stop, s, s_0, count_n, u_0, u_n, max_n \ | \ n \geq 1 \} \  \cup \\
    &\cup \{sq_{k,l}, sl_{k}, sl_{1,l}, sl_{2,l} \ | \ k \in \mathcal{N}, l \in \mathcal{D} \} 
\end{align*}

The set of membrane labels is given by: 

$H = \{1, INIT, COMP, OUTPUT\} \ \cup \ \{Q_{k,l} \ | \ k \in \mathcal{N}, l \in \mathcal{D} \} \ \cup \ \{LAMB_k \ | \ k \in \mathcal{N} \} \ \cup \ \{LAMB1_l \ | \ l \in \mathcal{D} \} \ \cup \ \{LAMB2_l \ | \ l \in \mathcal{D} \} \ \cup \ \{REDUCE_{0,k,l} \ | \ k \in \mathcal{N}, l \in \mathcal{D} \} \ \cup \ \{REDUCE_{1,k} \ | \ k \in \mathcal{N} \} \ \cup \ \{REDUCE_{2,l} \ | \ l \in \mathcal{D} \} \ \cup \ \{REDUCE_{3,l} \ | \ l \in \mathcal{D} \}$. 


The membrane structure $\mu$ can be defined as follows: 
    \begin{itemize}
        \item The skin membrane with label $1$, inside of which there exist:
        \begin{enumerate}[label*=\arabic*.]
            \item One membrane with label $INIT$.
            \item One membrane with label $OUTPUT$.
            \item One membrane with label $COMP$.
            \item One membrane with label $Q_{k,l} \ \forall k \in \mathcal{N}; \forall l \in \mathcal{D}$. Inside of each membrane $Q_{k,l}$:
            \begin{itemize}
                \item One membrane with label $REDUCE_{0,k,l} \ \forall k \in \mathcal{N}; \forall l \in \mathcal{D}$
            \end{itemize}
            \item One membrane with label $LAMB_k \ \forall k \in \mathcal{N}$. Inside of each membrane $LAMB_k$:
            \begin{itemize}
                \item One membrane with label $REDUCE_{1,k} \ \forall k \in \mathcal{N}$
            \end{itemize}
            \item One membrane with label $LAMB1_l \ \forall l \in \mathcal{D}$. Inside of each membrane $LAMB1_l$:
            \begin{itemize}
                \item One membrane with label $REDUCE_{2,l} \ \forall l \in \mathcal{D}$
            \end{itemize}
            \item One membrane with label $LAMB2_l \ \forall l \in \mathcal{D}$. Inside of each membrane $LAMB2_l$:
            \begin{itemize}
                \item One membrane with label $REDUCE_{3,l} \ \forall l \in \mathcal{D}$
            \end{itemize}
            
        \end{enumerate}
    \end{itemize} 

The initial multisets are $w_{0} = {y_0}$ and $w_{INIT} = x_{k,l}^{P}$ $\forall k \in \mathcal{N}, \ \forall l \in \mathcal{D}$. For any other membrane $h$, the initial multiset is $w_h = \emptyset$.

The set of rules $\mathcal{R}$ is given by the following rules, separated by stage, where $\lambda$ represents the empty multiset, the rule $RS_{i,j}$ represents the $j$-th rule of the $i$-th stage, the rules are defined $\forall k \in \mathcal{N}, \ \forall l \in \mathcal{D}$, the membranes $REDUCE_{0,k,l}, REDUCE_{1,k}, REDUCE_{2,l}, REDUCE_{3,l}$ are represented by $REDUCE$, and $\rho_{i,j}$ represents the priority of the rule $RS_{i,j}$: \\

\textbf{Stage 1 (Initialization)} \\
        $RS_{1,1} \equiv [x_{k,l}]^0_{INIT} \rightarrow x_{k,l}, xt_{k,l}, xl_{0,k,l}, xl_{1,k,l}, xl_{2,k,l} [ \ ]^0_{INIT} $ \\
        $RS_{1,2} \equiv [la_{k}]^0_{INIT} \rightarrow laq_{0,k,l}, la_{0,k} [ \ ]^0_{INIT} $ \\
        $RS_{1,3} \equiv [la_{1,l}]^0_{INIT} \rightarrow laq_{1,k,l}, la_{1,l} [ \ ]^0_{INIT} $ \\
        $RS_{1,4} \equiv [la_{2,l}]^0_{INIT} \rightarrow laq_{2,k,l}, la_{2,l} [ \ ]^0_{INIT} $ \\
        $RS_{1,5} \equiv [y_0 \rightarrow y_{0,k,l}, yl_{k}, yl_{1,l}, yl_{2,l}]^0_0$ \\ 
        $RS_{1,6} \equiv y_{0,k,l} [\ ]^0_{Q_{k,l}} \rightarrow [y_{0}, p_0^{\kappa_0}, ct_0^{\kappa_1}]^-_{Q_{k,l}} $
        with $\kappa_0 = \lfloor \frac{P \cdot \omega_k \gamma_{k,l}}{\beta_k} \rfloor$, $\kappa_1 = \lfloor \frac{2 \cdot P \cdot a_{kl} b_{kl}}{\beta_k} \rfloor$ and $a_{kl}, b_{kl}$ given by $\frac{\partial c_{kl}(q_{kl}^t)}{\partial q_{kl}} = 2 a^2_{kl} q_{kl} + 2 a_{kl} b_{kl}$   \\
        $RS_{1,7} \equiv yl_{k} [ \ ]^0_{LAMB_{k}} \rightarrow [y_{0}, n_0^{\lfloor s_k \cdot P \rfloor}  ]^-_{LAMB_{k}} $ \\
        $RS_{1,8} \equiv yl_{1,l} [ \ ]^0_{LAMB1_l} \rightarrow [y_{0}, p_0^{\lfloor \underline{d}_l \cdot P \rfloor}  ]^-_{LAMB1_l} $ \\
        $RS_{1,9} \equiv yl_{2,l} [ \ ]^0_{LAMB2_l} \rightarrow [y_{0}, n_0^{\lfloor \overline{d}_l \cdot P \rfloor}  ]^-_{LAMB2_l} $ \\
        $RS_{1,10} \equiv x_{k,l} [\ ]^-_{Q_{k,l}} \rightarrow [p, c_0]^-_{Q_{k,l}}$ \\
        $RS_{1,11} \equiv laq_{0,k,l} [\ ]^-_{Q_{k,l}} \rightarrow [n_0]^-_{Q_{k,l}}$ \\
        $RS_{1,12} \equiv laq_{1,k,l} [\ ]^-_{Q_{k,l}} \rightarrow [p_0]^-_{Q_{k,l}} $ \\
        $RS_{1,13} \equiv laq_{2,k,l} [\ ]^-_{Q_{k,l}} \rightarrow [n_0]^-_{Q_{k,l}}$ \\
        $RS_{1,14} \equiv xl_{0,k,l} [\ ]^-_{LAMB_k} \rightarrow [p_0]^-_{LAMB_k} $ \\
        $RS_{1,15} \equiv la_{0,k} [\ ]^-_{LAMB_k} \rightarrow [p]^-_{LAMB_k} $ \\
        $RS_{1,16} \equiv xl_{1,k,l} [\ ]^-_{LAMB1_l} \rightarrow [n_0]^-_{LAMB1_l} $ \\
        $RS_{1,17} \equiv la_{1,l} [\ ]^-_{LAMB1_l} \rightarrow [p]^-_{LAMB1_l} $ \\
        $RS_{1,18} \equiv xl_{2,k,l} [\ ]^-_{LAMB2_l} \rightarrow [p_0]^-_{LAMB2_l} $ \\
        $RS_{1,19} \equiv la_{2,l} [\ ]^-_{LAMB2_l} \rightarrow [p]^-_{LAMB2_l}$ \\

    \textbf{Stage 2 (Update)} 
        \begin{itemize}
            \item Update of $q_{kl}^t$ 
        \end{itemize}
        $RS_{2,1} \equiv [ct_0 \rightarrow ct_1]^-_{Q_{kl}}$ \\
        $RS_{2,2} \equiv [y_0 \rightarrow y_1]^-_{Q_{kl}}$ \\
        $RS_{2,3} \equiv [c_0 \rightarrow c_1^{\lfloor 2 \cdot P \cdot a_{kl}^2 \rfloor }]^-_{Q_{kl}}$
        with $a_{kl}$ given by $\frac{\partial c_{kl}(q_{kl}^t)}{\partial q_{kl}} = 2 a^2_{kl} q_{kl} + 2 a_{kl} b_{kl}$ \\
        $RS_{2,4} \equiv [ct_1 \rightarrow ct_2]^-_{Q_{kl}}$ \\
        $RS_{2,5} \equiv [y_1 \rightarrow y_2]^-_{Q_{kl}}$ \\
        $RS_{2,6} \equiv [c_1^{\lfloor P \cdot \beta^k \rfloor } \rightarrow n_0]^-_{Q_{kl}}$ \\
        $RS_{2,7} \equiv [c_1^{\lceil P \cdot \beta^k / 2 \rceil } \rightarrow n_0]^-_{Q_{kl}}$ with $\rho_{2,6} > \rho_{2,7}$ \\
        $RS_{2,8} \equiv [c_1 \rightarrow \lambda]^-_{Q_{kl}}$ with $\rho_{2,7} > \rho_{2,8}$ \\
        $RS_{2,9} \equiv [ct_2 \rightarrow n_0]^-_{Q_{kl}}$ \\
        $RS_{2,10} \equiv [y_2 \rightarrow y_3]^-_{Q_{kl}}$ \\
        $RS_{2,11} \equiv y_3 [ \ ]^0_{REDUCE} \rightarrow [ y_4 ]^-_{REDUCE}$ \\
        $RS_{2,12} \equiv p_0 [ \ ]^-_{REDUCE} \rightarrow [ p_0 ]^-_{REDUCE}$ \\
        $RS_{2,13} \equiv n_0 [ \ ]^-_{REDUCE} \rightarrow [ n_0 ]^-_{REDUCE}$ \\
        $RS_{2,14} \equiv [y_4 \rightarrow y_5]^-_{REDUCE}$ \\
        $RS_{2,15} \equiv [p_0, n_0 \rightarrow \lambda]^-_{REDUCE}$ \\
        $RS_{2,16} \equiv [p_0 \rightarrow p]^-_{REDUCE}$ with $\rho_{2,15} > \rho_{2,16}$ \\
        $RS_{2,17} \equiv [n_0 \rightarrow n]^-_{REDUCE}$ with $\rho_{2,15} > \rho_{2,17}$ \\
        $RS_{2,18} \equiv [p^2 \rightarrow p]^-_{REDUCE}$ \\
        $RS_{2,19} \equiv [p \rightarrow \lambda]^-_{REDUCE}$ with $\rho_{2,18} > \rho_{2,19}$ \\
        $RS_{2,20} \equiv [n^2 \rightarrow n]^-_{REDUCE}$ \\
        $RS_{2,21} \equiv [n \rightarrow \lambda]^-_{REDUCE}$ with $\rho_{2,20} > \rho_{2,21}$ \\
        $RS_{2,22} \equiv [s, y_5 \rightarrow s_0, y_5]^-_{REDUCE}$ with $\rho_{2,16} > \rho_{2,22}$, $\rho_{2,17} > \rho_{2,22}$, $\rho_{2,19} > \rho_{2,22}$, and $\rho_{2,21} > \rho_{2,22}$ \\
        $RS_{2,23} \equiv [y_5]^-_{REDUCE} \rightarrow y_6 [ \ ]^+_{REDUCE}$ with $\rho_{2,22} > \rho_{2,23}$  \\
        $RS_{2,24} \equiv [ p^{10} ]^+_{REDUCE} \rightarrow p \ [ \ ]^+_{REDUCE}$ \\
        $RS_{2,25} \equiv [ p^{5} ]^+_{REDUCE} \rightarrow p \ [ \ ]^+_{REDUCE}$ with $\rho_{2,24} > \rho_{2,25}$   \\
        $RS_{2,26} \equiv [ n^{10} ]^+_{REDUCE} \rightarrow n \ [ \ ]^+_{REDUCE}$ \\
        $RS_{2,27} \equiv [ n^{5} ]^+_{REDUCE} \rightarrow n \ [ \ ]^+_{REDUCE}$ with $\rho_{2,26} > \rho_{2,27}$   \\
        $RS_{2,28} \equiv y_6 \ [ \ ]^+_{REDUCE} \rightarrow y_7 \  [ rem ]^0_{REDUCE}$ with $\rho_{2,25} > \rho_{2,28}$ and $\rho_{2,27} > \rho_{2,28}$  \\
        $RS_{2,29} \equiv [ p  \rightarrow \lambda ]^0_{REDUCE}$ \\
        $RS_{2,30} \equiv [ n  \rightarrow \lambda ]^0_{REDUCE}$ \\
        $RS_{2,31} \equiv [ s_0  \rightarrow s ]^0_{REDUCE}$ \\ 
        $RS_{2,32} \equiv  [ y_7 ]^-_{Q_{kl}} \rightarrow y_{8,k,l} \  [ \ ]^+_{Q_{kl}}$ \\
        $RS_{2,33} \equiv [p, n \rightarrow \lambda]^+_{Q_{kl}}$ \\
        $RS_{2,34} \equiv [p \rightarrow o]^+_{Q_{kl}}$ with $\rho_{2,33} > \rho_{2,34}$ \\
        $RS_{2,35} \equiv [n \rightarrow \lambda]^+_{Q_{kl}}$ with $\rho_{2,33} > \rho_{2,35}$ \\
        $RS_{2,36} \equiv [y_{8,k,l} \rightarrow y_{9,k,l} ]^0_0$ \\
        $RS_{2,37} \equiv  [ o ]^+_{Q_{kl}} \rightarrow o_{0,k,l}, i_{k,l}, xt_{1,k,l} \  [ \ ]^+_{Q_{kl}}$ \\
        $RS_{2,38} \equiv  y_{9,k,l} \ [ \ ]^+_{Q_{kl}} \rightarrow y_{10} \ [ rem ]^0_{Q_{kl}}$ with $\rho_{2,37} > \rho_{2,38}$ 
        
        \begin{itemize}
            \item Update of $\lambda_{k}^t$ 
        \end{itemize}
        $RS_{2,39} \equiv [y_0 \rightarrow y_1]^-_{LAMB_k}$ \\
        $RS_{2,40} \equiv [y_1 \rightarrow y_2]^-_{LAMB_k}$ \\
        $RS_{2,41} \equiv [y_2 \rightarrow y_3]^-_{LAMB_k}$ \\
        $RS_{2,42} \equiv  [ y_7 ]^-_{LAMB_k} \rightarrow yla_{8,k} \  [ \ ]^+_{LAMB_k}$ \\
        $RS_{2,43} \equiv [p, n \rightarrow \lambda]^+_{LAMB_k}$ \\
        $RS_{2,44} \equiv [p \rightarrow o]^+_{LAMB_k}$ with $\rho_{2,43} > \rho_{2,44}$ \\
        $RS_{2,45} \equiv [n \rightarrow \lambda]^+_{LAMB_k}$ with $\rho_{2,43} > \rho_{2,45}$ \\
        $RS_{2,46} \equiv [yla_{8,k} \rightarrow yla_{9,k} ]^0_0$ \\
        $RS_{2,47} \equiv  [ o ]^+_{LAMB_k} \rightarrow lao_{0,k} \  [ \ ]^+_{LAMB_k}$ \\
        $RS_{2,48} \equiv  yla_{9,k} \ [ \ ]^+_{LAMB_k} \rightarrow \ [ rem ]^0_{LAMB_k}$ with $\rho_{2,47} > \rho_{2,48}$ 

        \begin{itemize}
            \item Update of $\lambda_{k}^{1^t}$ 
        \end{itemize}
        $RS_{2,48} \equiv [y_0 \rightarrow y_1]^-_{LAMB1_l}$ \\
        $RS_{2,49} \equiv [y_1 \rightarrow y_2]^-_{LAMB1_l}$ \\
        $RS_{2,50} \equiv [y_2 \rightarrow y_3]^-_{LAMB1_l}$ \\
        $RS_{2,51} \equiv  [ y_7 ]^-_{LAMB1_l} \rightarrow yla_{1,8,l} \  [ \ ]^+_{LAMB1_l}$ \\
        $RS_{2,52} \equiv [p, n \rightarrow \lambda]^+_{LAMB1_l}$ \\
        $RS_{2,53} \equiv [p \rightarrow o]^+_{LAMB1_l}$ with $\rho_{2,52} > \rho_{2,53}$ \\
        $RS_{2,54} \equiv [n \rightarrow \lambda]^+_{LAMB1_l}$ with $\rho_{2,52} > \rho_{2,54}$ \\
        $RS_{2,55} \equiv [yla_{1,8,l} \rightarrow yla_{1,9,l} ]^0_0$ \\
        $RS_{2,56} \equiv  [ o ]^+_{LAMB1_l} \rightarrow lao_{1,l} \  [ \ ]^+_{LAMB1_l}$ \\
        $RS_{2,57} \equiv  yla_{1,9,l} \ [ \ ]^+_{LAMB1_l} \rightarrow \ [ rem ]^0_{LAMB1_l}$ with $\rho_{2,56} > \rho_{2,57}$ 

        \begin{itemize}
            \item Update of $\lambda_{k}^{2^t}$ 
        \end{itemize}
        $RS_{2,58} \equiv [y_0 \rightarrow y_1]^-_{LAMB2_l}$ \\
        $RS_{2,59} \equiv [y_1 \rightarrow y_2]^-_{LAMB2_l}$ \\
        $RS_{2,60} \equiv [y_2 \rightarrow y_3]^-_{LAMB2_l}$ \\
        $RS_{2,61} \equiv  [ y_7 ]^-_{LAMB2_l} \rightarrow yla_{2,8,l} \  [ \ ]^+_{LAMB2_l}$ \\
        $RS_{2,62} \equiv [p, n \rightarrow \lambda]^+_{LAMB2_l}$ \\
        $RS_{2,63} \equiv [p \rightarrow o]^+_{LAMB2_l}$ with $\rho_{2,62} > \rho_{2,63}$ \\
        $RS_{2,64} \equiv [n \rightarrow \lambda]^+_{LAMB2_l}$ with $\rho_{2,62} > \rho_{2,64}$ \\
        $RS_{2,65} \equiv [yla_{2,8,l} \rightarrow yla_{2,9,l} ]^0_0$ \\
        $RS_{2,66} \equiv  [ o ]^+_{LAMB2_l} \rightarrow lao_{2,l} \  [ \ ]^+_{LAMB2_l}$ \\
        $RS_{2,67} \equiv  yla_{2,9,l} \ [ \ ]^+_{LAMB2_l} \rightarrow \ [ rem ]^0_{LAMB2_l}$ with $\rho_{2,66} > \rho_{2,67}$ \\

        \begin{itemize}
            \item Update of sequence terms $a_t$
        \end{itemize}
        $RS_{2,68} \equiv [count_0^{1024} \rightarrow u_0, s]^0_{INIT}$ \\
        $RS_{2,69} \equiv [s]^0_{INIT} \rightarrow sq_{k,l}, sl_{k}, sl_{1,l}, sl_{2,l} \ [ \ ]^0_{INIT}$ \\
        $RS_{2,70} \equiv sq_{k,l} [\ ]^-_{Q_{k,l}} \rightarrow [s]^-_{Q_{k,l}}$ \\
        $RS_{2,71} \equiv sl_{k} [\ ]^-_{LAMB_k} \rightarrow [s]^-_{LAMB_k} $ \\
        $RS_{2,72} \equiv sl_{1,l} [\ ]^-_{LAMB1_l} \rightarrow [s]^-_{LAMB1_l} $ \\
        $RS_{2,73} \equiv sl_{2,l} [\ ]^-_{LAMB2_l} \rightarrow [s]^-_{LAMB2_l} $ \\
        $RS_{2,74} \equiv s [\ ]^0_{REDUCE} \rightarrow [s]^0_{REDUCE} $ \\
        $RS_{2,75} \equiv [u_0^{10} \rightarrow max_0]^0_{INIT}$ \\
        $RS_{2,76} \equiv [max_n, count_0 \rightarrow max_n, count_{n+1}]^0_{INIT}$ for $n \geq 1$ \\
        $RS_{2,77} \equiv [count_n^{(2^{10+n})} \rightarrow u_n, s]^0_{INIT}$ for $n \geq 1$ \\
        $RS_{2,78} \equiv [u_n, max_{n-1} \rightarrow max_n]^0_{INIT}$ for $n \geq 1$ \\

    \textbf{Stage 3 (Comparison)} \\
        $RS_{3,1} \equiv y_{10}^{m \cdot n} \ [ \ ]^0_{COMP} \rightarrow [ y_{11} ]^-_{COMP} $ \\ 
        $RS_{3,2} \equiv o_{0,k,l} \ [ \ ]^-_{COMP} \rightarrow [ o_{1,k,l} ]^-_{COMP} $ \\ 
        $RS_{3,3} \equiv xt_{k,l} \ [ \ ]^-_{COMP} \rightarrow [ xt_{k,l} ]^-_{COMP} $ \\
        $RS_{3,4} \equiv xt_{1,k,l} \ [ \ ]^-_{COMP} \rightarrow [ xt_{1,k,l} ]^-_{COMP} $ \\
        $RS_{3,5} \equiv  [ y_{11} ]^-_{COMP} \rightarrow rem \  [ y_{12} ]^0_{COMP}$ with $\rho_{3,2} > \rho_{3,5}$, $\rho_{3,3} > \rho_{3,5}$ and $\rho_{3,4} > \rho_{3,5}$ \\
        $RS_{3,6} \equiv  [ o_{1,k,l} ]^0_{COMP} \rightarrow o_{2,k,l} \  [ \ ]^0_{COMP}$  \\
        $RS_{3,7} \equiv [xt_{k,l}, xt_{1,k,l} \rightarrow \lambda]^0_{COMP}$ \\
        $RS_{3,8} \equiv [xt_{k,l}, y_{12} \rightarrow y_{13}]^0_{COMP}$ with $\rho_{3,7} > \rho_{3,8}$ \\
        $RS_{3,9} \equiv [xt_{1,k,l}, y_{12} \rightarrow y_{13}]^0_{COMP}$ with $\rho_{3,7} > \rho_{3,9}$ \\
        $RS_{3,10} \equiv [xt_{k,l} \rightarrow \lambda]^0_{COMP}$ with $\rho_{3,8} > \rho_{3,10}$ \\
        $RS_{3,11} \equiv [xt_{1,k,l} \rightarrow \lambda]^0_{COMP}$ with $\rho_{3,9} > \rho_{3,11}$ \\
        $RS_{3,12} \equiv [y_{12} \rightarrow stop]^0_{COMP}$ with $\rho_{3,8} > \rho_{3,12}$ and $\rho_{3,9} > \rho_{3,12}$ \\
        $RS_{3,13} \equiv  [ o_{2,k,l}  \rightarrow o_{3,k,l} ]^0_0$  \\
        $RS_{3,14} \equiv  [ stop ]^0_{COMP} \rightarrow stop \  [ \ ]^0_{COMP}$  \\
        $RS_{3,15} \equiv  [ y_{13} ]^0_{COMP} \rightarrow y_{14} \  [ \ ]^0_{COMP}$  \\
        $RS_{3,16} \equiv  [ o_{3,k,l}  \rightarrow o_{4,k,l} ]^0_0$  \\
        $RS_{3,17} \equiv stop \ [ \ ]^0_{OUTPUT} \rightarrow [ stop ]^-_{OUTPUT}$ \\
        $RS_{3,18} \equiv  y_{14} \ [ \ ]^0_{INIT} \rightarrow [ y_{15} ]^-_{INIT} $ \\ 
        $RS_{3,19} \equiv   o_{4,k,l} \ [ \ ]^-_{OUTPUT} \rightarrow [ o_{k,l} ]^-_{OUTPUT} $  \\
        $RS_{3,20} \equiv  [ o_{4,k,l}  \rightarrow \lambda ]^0_0$ with $\rho_{3,19} > \rho_{3,20}$  \\
        $RS_{3,21} \equiv   [ stop ]^-_{OUTPUT} \rightarrow rem \ [ \ ]^0_{OUTPUT} $ with $\rho_{3,19} > \rho_{3,21}$ \\
        $RS_{3,22} \equiv  i_{k,l} \ [ \ ]^-_{INIT} \rightarrow [ x_{k,l} ]^-_{INIT} $ \\
        $RS_{3,23} \equiv  lao_{0,k} \ [ \ ]^-_{INIT} \rightarrow [ la_{k} ]^-_{INIT} $ \\
        $RS_{3,24} \equiv  lao_{1,l} \ [ \ ]^-_{INIT} \rightarrow [ la_{1,l} ]^-_{INIT} $ \\
        $RS_{3,25} \equiv  lao_{2,l} \ [ \ ]^-_{INIT} \rightarrow [ la_{2,l} ]^-_{INIT} $ \\
        $RS_{3,26} \equiv [ y_{15} ]^-_{INIT} \rightarrow y_{0} \ [ count_0 ]^0_{INIT}$ with $\rho_{3,22} > \rho_{3,26}$, $\rho_{3,23} > \rho_{3,26}$, $\rho_{3,24} > \rho_{3,26}$ and $\rho_{3,25} > \rho_{3,26}$ \\

    \textbf{Cleaning} \\
        $RS_{4,1} \equiv [rem \rightarrow \lambda]^c_h$ for each membrane $h \in H$ and for each charge $c \in EC$.




\end{appendices}


\bibliography{sn-bibliography}

\end{document}